\documentclass{article}

\usepackage{arxiv}

\usepackage[utf8]{inputenc} 
\usepackage[T1]{fontenc}    
\usepackage{hyperref}       
\usepackage{url}            
\usepackage{booktabs}       
\usepackage{amsfonts}       
\usepackage{nicefrac}       
\usepackage{microtype}      
\usepackage{lipsum}		
\usepackage{graphicx}
\usepackage{natbib}
\usepackage{doi}
\usepackage{multicol,multirow}
\usepackage{subcaption}
\usepackage{arydshln}
\usepackage{chngcntr}
\usepackage{multibib}
\usepackage{subfiles}
\usepackage{pifont}
\usepackage{amsmath}
\usepackage{tablefootnote}
\usepackage{authblk}

\newcites{SM}{Supplementary Material References} 

\newcommand{\cmark}{\ding{51}}%
\newcommand{\xmark}{\ding{55}}%

\title{Understanding cirrus clouds using explainable machine learning}

\date{} 					

\author[1]{\textbf{Kai Jeggle}}
\author[1]{\textbf{David Neubauer}}
\author[2]{\textbf{Gustau Camps-Valls}}
\author[1]{\textbf{Ulrike Lohmann}}

\affil[ ]{Corresponding author: \textit {kai.jeggle@env.ethz.ch}}
\affil[1]{Institute of Atmospheric and Climate Science,	ETH Zurich,  Zurich, Switzerland}
\affil[2]{Image Processing Laboratory, Universitat de València, València, Spain}

\date{}

\renewcommand{\shorttitle}{This work has been submitted to \textit{Environmental Data Science}}

\hypersetup{
pdftitle={Understanding cirrus clouds using explainable machine learning},
pdfsubject={physics.ao-ph, cs.LG},
pdfauthor={Kai Jeggle, David Neubauer, Gustau Camps-Valls, Ulrike Lohmann},
pdfkeywords={Cirrus, Machine Learning, Explainable AI, Explainable Machine Learning, XAI, SHAP, Shapley values},
}

\begin{document}
\maketitle

\begin{abstract}
    Cirrus clouds are key modulators of Earth's climate. Their dependencies on meteorological and aerosol conditions are among the largest uncertainties in global climate models. This work uses three years of satellite and reanalysis data to study the link between cirrus drivers and cloud properties. We use a gradient-boosted machine learning model and a Long Short-Term Memory (LSTM) network with an attention layer to predict the ice water content and ice crystal number concentration. The models show that meteorological and aerosol conditions can predict cirrus properties with $R^2 = 0.49$. Feature attributions are calculated with SHapley Additive exPlanations (SHAP) to quantify the link between meteorological and aerosol conditions and cirrus properties. For instance, the minimum concentration of supermicron-sized dust particles required to cause a decrease in ice crystal number concentration predictions is $2 \times 10^{-4}$ mg m\textsuperscript{-3}. The last 15 hours before the observation predict all cirrus properties.
\end{abstract}


\section{Introduction}

Cirrus clouds consist of ice crystals and occur at temperatures below -38°C in the upper troposphere \citep{sassen_global_2008}. Like all other clouds, they influence the Earth's radiative budget via the reflection of solar radiation and absorption and emission of terrestrial infrared radiation into space \citep{liou_influence_1986}. Depending on the cloud microphysical properties (CMP), namely ice water content $IWC$ and ice crystal number concentration $N_i$, the cloud radiative effect of cirrus varies significantly and ranges from a cooling effect at lower altitudes for optically thick cirrus to a warming effect at high altitudes for optically thin cirrus \citep{heymsfield_cirrus_2017}. However, due to the high spatiotemporal variability of cirrus, different formation pathways, and the non-linear dependence on environmental conditions and aerosols, constraining CMP and hence the cloud radiative effect remains an active field of study \citep{kramer_microphysics_2020, gasparini_cirrus_2018, gryspeerdt_ice_2018}. An improved process understanding of cirrus CMP will help to reduce uncertainties in global climate models and climate change projections and assess climate intervention methods targeted at cirrus clouds.
This study aims to quantify the impacts of individual cirrus drivers on the CMP. We approach this task with a two-stage process. First, we train a black-box machine learning model (ML) to predict $IWC$ and $N_i$, given the input of meteorological variables and aerosol concentrations. Next, we calculate the individual contributions to the predictions for each input feature using post-hoc feature attribution methods from the emerging field of eXplainable AI (XAI) \citep{belle_principles_2020}. Using XAI methods to understand and discover new physical laws via ML has been discussed in recent publications and is
increasingly adopted in the atmospheric science community \citep{ebert-uphoff_evaluation_2020, mcgovern_making_2019}. We train and explain a gradient-boosted regression tree on instantaneous data and a Long Short-Term Memory (LSTM) network \citep{hochreiter_long_1997} with an attention layer on time-resolved data. The former is aimed to yield better interpretability, while the latter incorporates temporal information of the predictors. 
The remainder of this paper is structured as follows: A brief background about cirrus drivers is presented in \ref{cirrus_background}, followed by a description of the data set, ML models, and feature attribution methods in section \ref{section2}. Section \ref{section3} presents the prediction performance of the ML models and the corresponding feature attributions. Finally, the results are summarized and discussed in section \ref{section4}.

\section{Cirrus driver}\label{cirrus_background}
Previous studies have found that cirrus CMP are mainly controlled by the ice crystal nucleation mode \citep{karcher_physically_2006}, ice origin \citep{kramer_microphysics_2020}, meteorological variables such as temperature and updraft speeds, and aerosol environment \citep{gryspeerdt_ice_2018}. There are two possible nucleation modes - homogeneous and heterogeneous nucleation. Homogeneous nucleation is the freezing of supercooled solution droplets below -38 °C and high ice supersaturations \citep{koop_water_2000}. It strongly depends on updraft velocity, producing higher $N_i$ with increasing updrafts \citep{jensen_susceptibility_2016}. Heterogeneous nucleation occurs via crystalline or solid aerosols acting as ice nucleating particles that can significantly lower the energy barrier needed for ice nucleation \citep{kanji_overview_2017}. Dust is considered the most important ice nucleating particle \citep{kanji_overview_2017} and is hence often the only ice nucleating particle considered in global climate models. The availability of ice nucleating particles can lead to fewer but larger crystals by suppressing homogeneous nucleation \citep{kuebbeler_dust_2014}. 

\citet{kramer_microphysics_2020} and \citet{wernli_trajectory-based_2016} showed that the history of the air parcel leading up to a cloud significantly influences cirrus CMP. Two main categories with respect to the ice origin were identified: Liquid-origin cirrus form by freezing cloud droplets, i.e., originating from mixed-phase clouds at temperatures above -38 °C and leading to higher IWC, in-situ cirrus form ice directly from the gas phase or freezing of solution droplets at temperatures below -38 °C and lead to thinner cirrus with lower IWC. Besides the ice origin, the air parcel's exposure to changing meteorological and aerosol conditions upstream of a cloud observation can impact the cirrus CMP. For instance, a high vertical velocity occurring prior to a cloud observation may increase $IWC$ and $N_i$ \citep{heymsfield_cirrus_2017}. In this study, we focus on the meteorological variables and aerosol environment as cirrus drivers. Nucleation mode and ice origin are classifications that are themselves controlled by the cirrus drivers.

\section[Methods and Data]{Methods and Data}\label{section2}

\subsection{Data}

This study combines satellite observations of cirrus clouds and reanalysis data of meteorology and aerosols to predict $IWC$ and $N_i$. We use the DARDAR-Nice \citep{sourdeval_ice_2018} product for satellite-based cirrus retrievals. The DARDAR-Nice product combines measurements from CloudSat's radar and CALIOP's lidar to obtain vertically resolved retrievals for $IWC$ and $N_{i}$ for ice crystals > 5 {\textmu}m in cirrus clouds. DARDAR-Nice also provides information about each layer's distance to the cloud top. Additionally, we calculate a cloud thickness feature, which  provides information about the vertical extent of a cloud. To represent cirrus cloud drivers, we use the ERA5 \citep{hersbach_era5_2018} and MERRA2 \citep{global_modeling_and_assimilation_office_merra-2_2015} reanalysis data sets for meteorological and aerosol drivers, respectively. All data sources are co-located on a 0.25°x0.25°x300m spatial and hourly temporal resolution corresponding to the original ERA5 resolution. Due to the narrow swath of the satellite overpass (i.e., low spatial coverage) and long revisiting time (16 days), studies about the temporal development of cirrus clouds and their temporal dependencies on driver variables are not possible given only instantaneous DARDAR-Nice data co-located with driver variables.

\begin{figure}
    \includegraphics[width=0.8\textwidth]{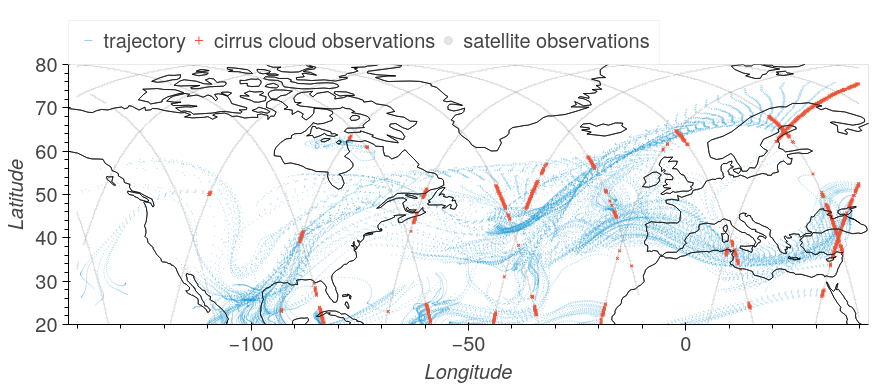}
    \centering
    \caption{Illustrative example of data for one hour. DARDAR-Nice satellite observations along the satellite overpass are displayed as grey dots. A vertical profile of 56 300 m thick vertical layers is available for each satellite observation. To keep the diagram uncluttered, only one vertical level is shown. Red crosses mark observations containing cirrus clouds. Blue dashes represent the corresponding $48$ hours of backtrajectories calculated for each cirrus observation. Along each trajectory, meteorological and aerosol variables are traced}
    \label{fig:sample_data}
\end{figure}

CMPs are not only determined by their instantaneous environmental variables but undergo a temporal development in which they are exposed to changing atmospheric states such as shifts in temperature, updraft velocities, and aerosol environment. Thus, we add a temporal dimension to our data set by calculating 48 hours of Lagrangian backtrajectories for every cirrus cloud observation in the DARDAR-Nice data set with LAGRANTO \citep{sprenger_lagranto_2015}. An individual trajectory is calculated for every cirrus layer in an atmospheric column. For instance, for a 900 m thick cloud consisting of three 300 m layers, three individual trajectories are calculated and initialized simultaneously and identical latitude/longitude but at different height levels. LAGRANTO uses ERA5 wind fields for the trajectory calculation. All meteorological and aerosol variables are traced along the trajectory. These variables represent the changing meteorology and aerosol environment along the air parcel upstream of a cirrus observation in the satellite data.
Figure \ref{fig:sample_data} exemplarily displays the satellite overpass (grey), cirrus cloud observations (red), and the corresponding backtrajectories (blue). 

To account for seasonal and regional co-variations, we include categorical variables representing the different seasons (\textit{DJF, MMA, JJA, SON}) and regions divided in 10° latitude bins into the data set. Additionally, the surface elevation of the ground below an air parcel is included to incorporate the influence of orographic lifting. Since an air parcel can travel through multiple 10° latitude bins in 48 hours and is exposed to varying topographies, both latitude bin and and surface elevation are also traced along the trajectory. Finally, we use the land water mask provided by the DARDAR-Nice dataset. Please refer to \cite{sourdeval_ice_2018} for more information. All used data variables are specified in Table \ref{table:data_description}. The last column in the table indicates whether the variable is available along the backtrajectories or just at the time of the satellite observation. We chose to study a domain ranging from 140°W to 40°E and 25°N to 80°N. This study area covers a wide range of climate zones and topographies. In the vertical direction, we consider $56$ vertical layers of 300 m thickness covering heights between 4080 m and 20580 m. Since we only consider cirrus clouds in this study, all data above -38 °C are omitted. The tropics are excluded from this study as convective regimes often drive liquid-origin cirrus clouds in this region, and convection is not well represented at this horizontal resolution. Due to gaps in the satellite data, we focus on the period between 2007-2009. Since sunlight can adversely affect the quality of satellite retrievals, we only consider nighttime observations.

\begin{table}
	\caption{Summary of data used in this study. Lagrangian backtrajectories were calculated 48 hours back in time for each cirrus observation. Variables with \cmark \ in the last column are available along backtrajectories (hourly data). Variables with \xmark \ are only available at the satellite overpass. Note: Dust concentrations are divided in particles smaller than 1 $\mu$ m \ ($DU_{sub}$) and particles larger than 1 $\mu$ m \ ($DU_{sup}$).}
	\centering
	\begin{tabular*}{\textwidth}{@{\extracolsep{\fill}}llcc@{}}\toprule%

         & Variable & Source & Backtrajectory \\\midrule
        \multirow{2}{*}{Target Variables} & Ice crystal number concentration ($N_i$) [mg m\textsuperscript{-3}] & \multirow{2}{*}{DARDAR-Nice} & \multirow{2}{*}{\xmark} \\
                         & Ice Water Content ($IWC$) [cm\textsuperscript{-3}] & & \\
        \hline
        \midrule
        
        \multirow{11}{*}{Predictor Variables}    & Temperature [K]                 & \multirow{3}{*}{ERA5}    & \multirow{3}{*}{\cmark}   \\
                              & Vertical velocity [Pa s\textsuperscript{-1}]       &               &               \\
                              & Horizontal wind speed [m s\textsuperscript{-1}]    &               &               \\
        
        \cmidrule{2-4}
                              & Dust aerosols\ ($DU_{sub}$, $DU_{sup}$) [mg kg\textsuperscript{-1}]     & \multirow{2}{*}{MERRA2}   & \multirow{2}{*}{\cmark}   \\ 
                              & Sulfate aerosols ($SO4$) [mg kg\textsuperscript{-1}]         &               &               \\
                             
        \cmidrule{2-4}
                              & Distance from cloud top [m]  & \multirow{3}{*}{DARDAR-Nice} & \multirow{3}{*}{\xmark}   \\
                              & Cloud thickness [m]          &                              &               \\
                              & Land water mask              &                              &               \\
        \cmidrule{2-4}
                              & Surface elevation [m]                       & \multirow{3}{*}{ERA5}             &  \cmark             \\
                              & Region [10° latitude bins]                  &                                   &   \cmark            \\
                              & Season                                      &                                   &   \xmark             \\
        
        \end{tabular*}%
	\label{table:data_description}
\end{table}

\subsection{Prediction Models}
 We aim to predict $IWC$ and $N_i$ of cirrus clouds with meteorological and aerosol variables, a regression task from an ML perspective. By adding latitude bin, season, and surface elevation to the input features, we account for regional, seasonal, and topographic variability. We approach the regression task with (1) instantaneous and (2) time-resolved input data. The former has the advantage of training the ML model on a tabular data set that best suits the explainability methods described in the next section. The latter data set enables the model to incorporate the temporal dependencies of cirrus drivers on the cirrus CMP but is more difficult to interpret. Note that only the input data has a temporal dimension, namely the backtrajectories.

\subsubsection{Instantaneous Prediction Model}
We train a gradient-boosted regression tree using the XGBoost algorithm \citep{chen_xgboost_2016}, which 
typically 
outperforms more complex deep learning models on tabular data  
\citep{shwartz-ziv_tabular_2021}. A separate XGBoost model is trained for each target variable.

\subsubsection{Time Resolved Prediction Model}

Given the temporally resolved backtrajectory data of cirrus drivers $x_t$ with $t \in [-48,0]$ (sequential features) and static features $x_{static}$, like cloud thickness and distance to cloud top, we predict the target variables $IWC$ and $N_i$ at $t = 0$,  i.e., the time of the satellite observation, using a single multi-output Long Short-Term Memory (LSTM) architecture \citep{hochreiter_long_1997}. From an ML perspective, this is a many-to-one regression problem. LSTMs are neural networks that capitalize on temporal dependencies in the input data. The LSTM encodes the sequential features into a latent representation called hidden state $h_t$ \eqref{eq1}.

\begin{align}
  & h_t = LSTM(x_t), t \in [-48,0] \label{eq1} \\
  & u_t = tanh(Wh_t + b) \label{eq2} \\
  & \alpha_t = \frac{exp(u_t u)}{\sum_t exp(u_t u)} \label{eq3} \\
  & z = \sum_t \alpha_t h_t \label{eq4}
\end{align}

\begin{figure}
	\centering
        \includegraphics[width=0.6\textwidth]{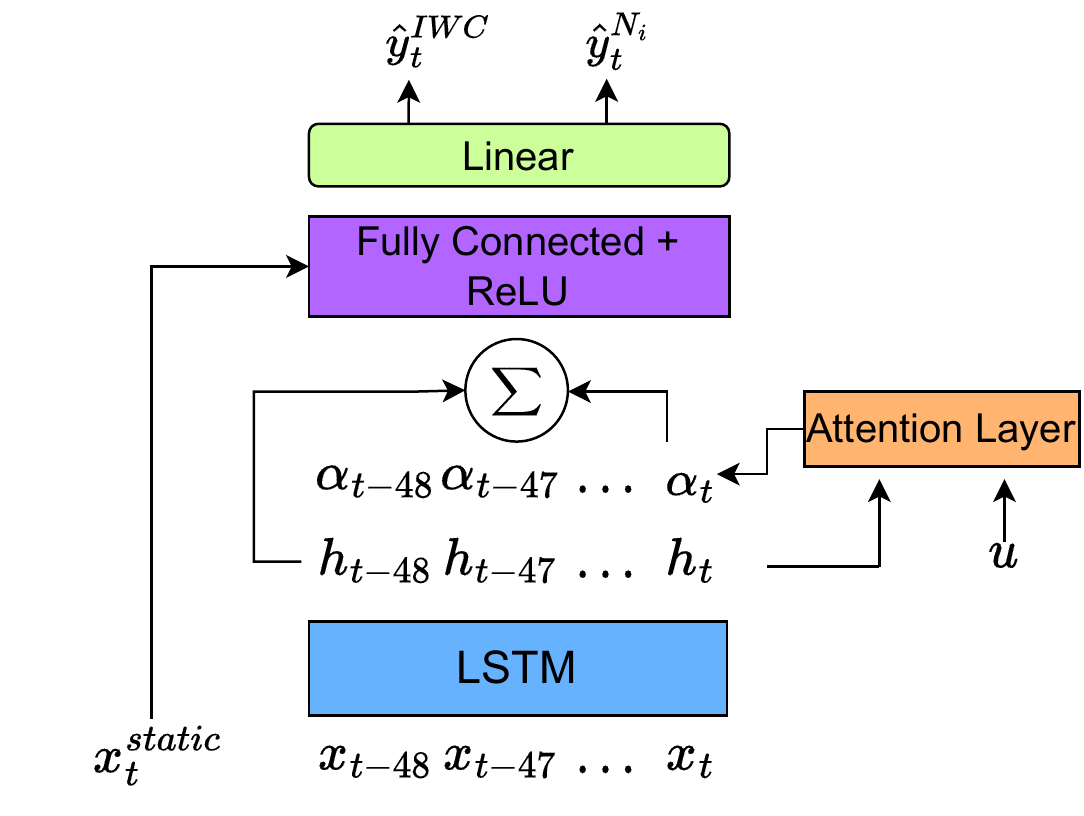}
	\caption{Architecture of the LSTM based model with integrated attention method to predict $IWC$ and 
                 $N_i$ of cirrus clouds with the temporal data set. Abbreviations: ReLU - Rectified Linear Unit}
        \label{figure:lstm_architecture}
\end{figure}

Then, the output of the LSTM is fed to a simple attention mechanism that determines the relative importance of each timestep in the sequential input data for the final prediction. The attention mechanism provides a built-in approach to interpreting the ML model's prediction for the temporal dimension. It comprises a single fully connected layer and a context vector $u$. First, a latent representation $u_t$ of $h_t$ is created by passing $h_t$ through the attention layer \eqref{eq2}, then the importance of each timestep $\alpha_t$ is calculated by multiplying $u_t$ with the context vector $u$. We normalize the importance weights $\alpha_t$ by applying the softmax function \eqref{eq3} so that all attention weights sum up to 1. Next, the weighted sum $z$ of the importance weights $\alpha_t$ and the LSTM hidden state $h_t$ is calculated \eqref{eq4}. The attention layer, as well as the context vector $u$, are randomly initialized and learned during the training through backpropagation together with the weights of the LSTM and other fully connected layers. To make the final prediction $y \in\mathbb{R}_+^2$, $z$ is concatenated with the static input features $x_{static}$ and fed through a set of fully connected layers. Figure \ref{figure:lstm_architecture} visualizes the whole architecture consisting of LSTM, attention layer, fully connected layers, and finally, a linear layer.

\subsubsection{Experimental Setup}
To train and evaluate the ML models, we split the data into 80\% training, 10\% validation, and 10\% test data. Each cloud layer containing cirrus clouds is considered an individual sample, i.e., a cloud consisting of 3 300m layers is considered as three different samples. 
To prevent overfitting due to spatiotemporally correlated samples in training and test sets, samples from the same month are put in the same split. In total, the data set consists of ~6 million  samples. 
Due to the large spread in scales, the target variables and aerosol concentrations are logarithmically transformed. The categorical features are one-hot encoded. For the XGBoost model, no further pre-processing is required. For the LSTM model, we standardize the continuous sequential and static features by removing the mean and dividing them by the standard deviation of the training samples. The hyperparameters of the models are tuned using Bayesian optimization. Fifty hyperparameter configurations are tested per model during the optimization. The final hyperparameters are displayed in Appendix \ref{appendixExperimental setup}.

\subsection{XAI}

We apply post-hoc feature attribution methods to the trained models. 
For each sample, the marginal contribution of each feature towards the prediction is calculated. By combining explanations for all samples, we can gain insight into the internal mechanics of the model and, eventually, an indication of the underlying physical processes.  We focus on the instantaneous predictive model 
 and only briefly interpret the temporal model by analyzing the attention weights. We use SHapley Additive exPlanations (SHAP) \citep{lundberg_unified_2017} using the TreeSHAP \citep{lundberg_local_2020} implementation to calculate feature attributions. SHAP is a widely applied XAI method based on game-theoretic Shapley values. For each feature, SHAP outputs the change in expected model prediction, i.e., if a feature contributes to an increase or decrease in the prediction.

To evaluate the quality of black-box ML model explanations, it is customary 
to calculate faithfulness and stability metrics. The notion of faithfulness states how well a given explanation represents the actual behavior of the ML model \citep{alvarez_melis_towards_2018}, and stability measures how robust an explanation is towards small changes in the input data \citep{alvarez_melis_towards_2018, agarwal_rethinking_2022}. In this study, we use \textit{Estimated Faithfulness} \citep{alvarez_melis_towards_2018}, which measures the correlation between the relative importance assigned to a feature by the explanation method and the effect of each feature on the performance of the prediction model. Higher importance should have a higher effect and vice versa. Explanation stability is evaluated using Relative Input Stability (RIS) and Relative Output Stability (ROS) \citep{agarwal_rethinking_2022}. Implementation details of the XAI evaluation metrics are described in Appendix \ref{appendixXAIMetrics}. To further validate our explanations, we compare them with LIME \citep{ribeiro_why_2016} and baseline with randomly generated attributions. LIME is another popular XAI method that calculates feature attributions by fitting linear models in the neighborhood of a single sample.

\section[Results]{Results}\label{section3}
We first present the predictive performance of the XGBoost and LSTM-based models. The feature attributions of the XGBoost model are discussed. Finally, we analyze the weights of the attention layer in the LSTM-based model to understand which timesteps are important for the prediction tasks.

\subsection{Cirrus Cloud Prediction Performance}
The prediction performance of the ML models and a linear baseline regression are displayed in Table \ref{predictive_performance}. Both ML models outperform the baseline, and the LSTM-based model performs best overall with 0.49 (0.40) $R^2$, and root mean squared error of 0.35 (0.33) for $IWC$ ($N_i$). This shows that the backtrajectories contain useful information for cirrus cloud prediction. Like the LSTM-based model, the XGBoost model can capture non-linear relationships in the data. Thus, it is fair to continue focusing on that model for model explainability.

\begin{table}
    \caption{Performance metrics evaluated on independent test data for the regression task of predicting cirrus CMP using meteorological and aerosol variables. Each model was trained $10$ times; the values in the table represent the mean and standard deviation over the $10$ model runs. Note: Performance metrics are calculated on log-transformed values of target variables. Abbreviations: RMSE - Root Mean Squared Error, $R^2$ - Coefficient of determination.}
    
    \centering
    \begin{tabular*}{\textwidth}{@{\extracolsep{\fill}}llcccc@{}}\toprule%
             & & \multicolumn{2}{@{}c@{}}{$IWC$}& \multicolumn{2}{@{}c@{}}{$N_i$}
             \\\cmidrule{3-4}\cmidrule{5-6}%
            Model & Input Data & RMSE & $R^2$ & RMSE & $R^2$\\\midrule
            Linear regression (baseline) & Instantaneous &0.44 &0.24 &0.45 &0.19\\
            XGBoost& Instantaneous &$0.40\pm0.01$ &$0.38\pm0.02$ &$0.42\pm0.01$ &$0.35\pm0.02$\\
            LSTM& Temporal & $0.35\pm0.03$ &$0.49\pm0.04$ &$0.33\pm0.02$ & $0.42\pm0.03$\\
        \end{tabular*}%
    \label{predictive_performance}
\end{table}

\subsection{Feature attributions with SHAP}


\begin{figure}
	\centering
	\includegraphics[width=0.9\textwidth]{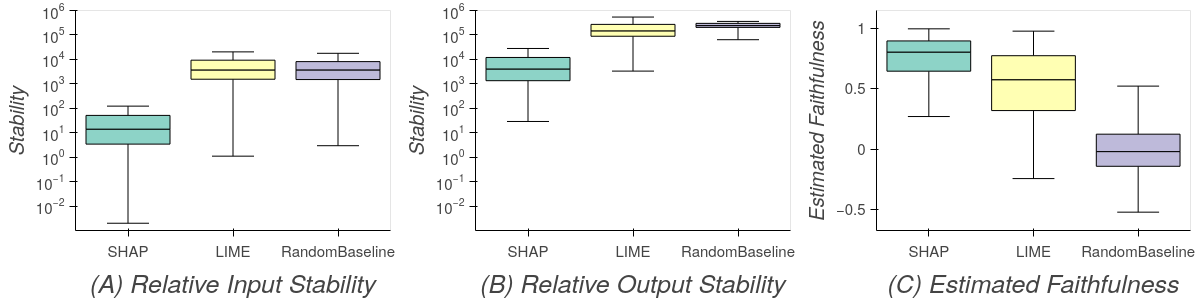}
        \caption{XAI evaluation metrics for SHAP, LIME, and a random baseline feature attribution of the XGBoost model. For the stability metrics RIS \textbf{(A)} and ROS \textbf{(B)}, lower values indicate more stable explanations, i.e., more robust toward small changes of the feature values. Estimated Faithfulness \textbf{(C)} indicates whether features with high importance attributed by the feature attribution method are important for the prediction performance, where 1 denotes perfect estimated faithfulness.}
        \label{fig:xai_evaluation}
\end{figure}

\begin{figure}
 \begin{subfigure}[t]{.48\textwidth}
  \centering
  \includegraphics[width=\linewidth]{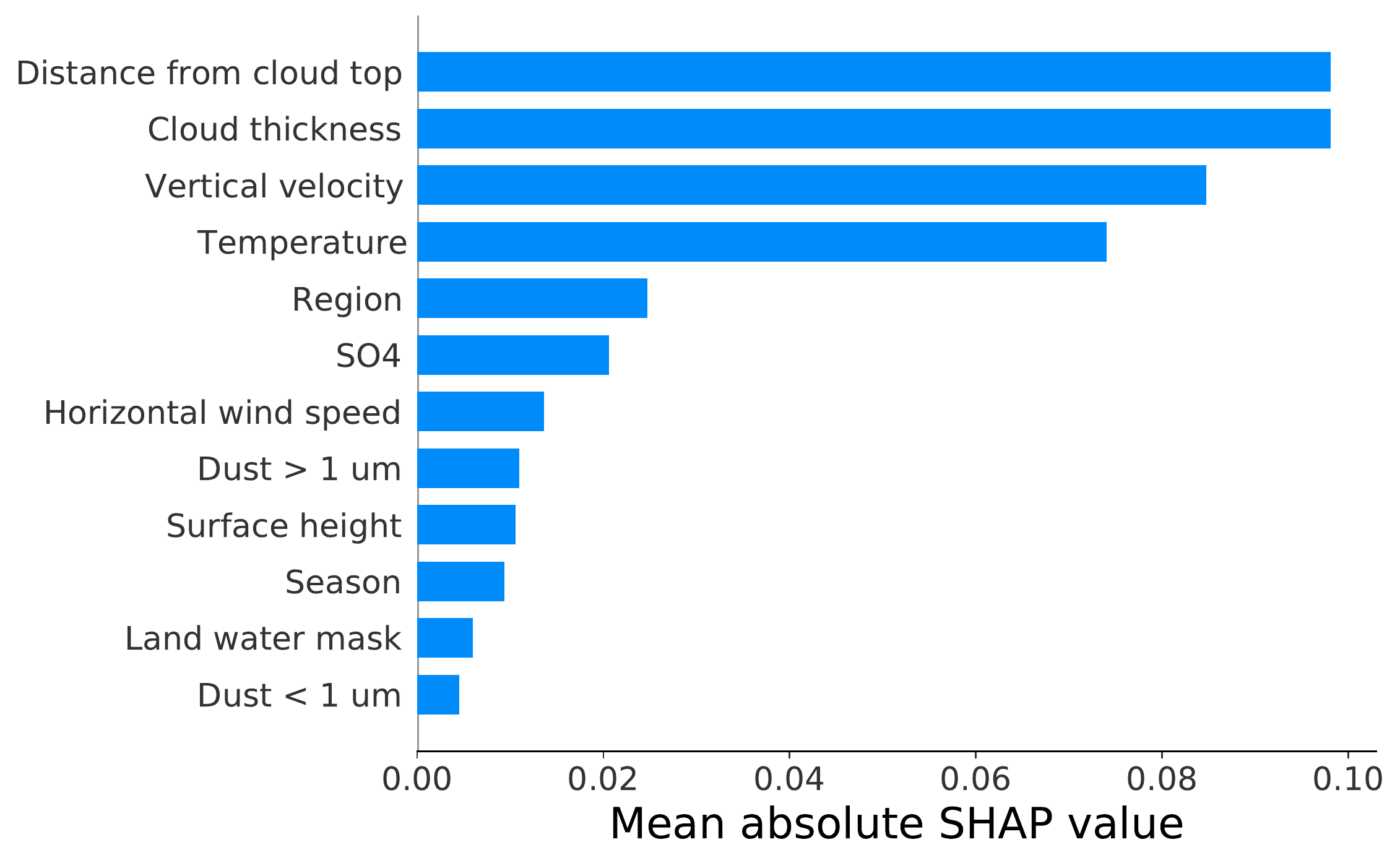}
  \caption{$IWC$}
 \end{subfigure}
 \hfill
 \begin{subfigure}[t]{.48\textwidth}
  \centering
  \includegraphics[width=\linewidth]{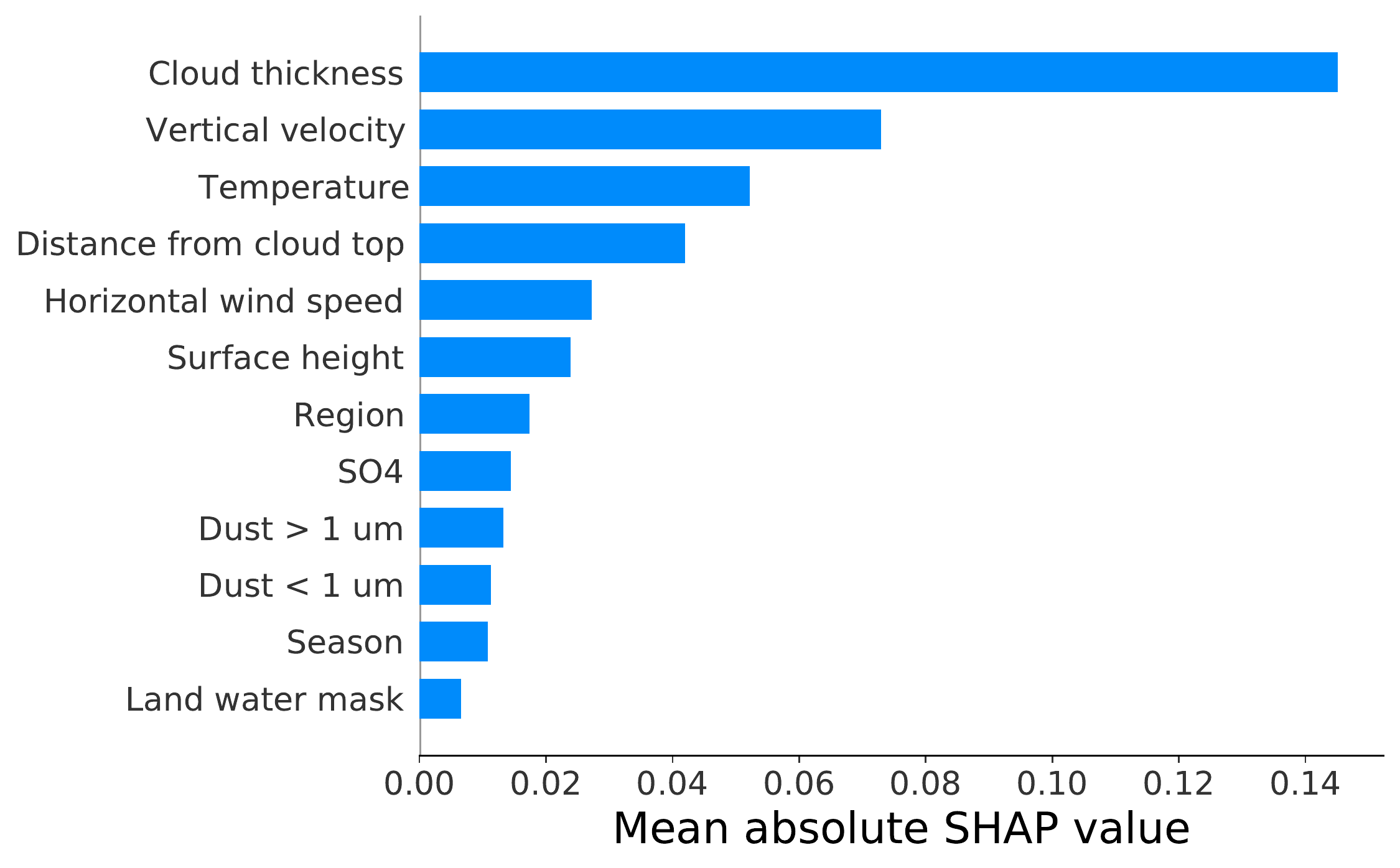}
  \caption{$N_i$}
 \end{subfigure}
{\caption{Mean absolute SHAP values of each feature for $IWC$ (\textbf{a}) and $N_i$ (\textbf{b}). The higher the value, the higher the contribution to the prediction, i.e., the more important the feature. }\label{fig:absolute_shap_values}}
\end{figure}

Figure \ref{fig:xai_evaluation} shows that SHAP explanations are more stable and have higher faithfulness than the other feature attribution methods and are thus best suited for our study. For each feature, the mean absolute SHAP value aggregated over all test samples is displayed in Figure \ref{fig:absolute_shap_values} for $IWC$ (left) and $N_i$ (right). It can be seen that the vertical extent of the cloud and the position of the cloud layer within the cloud are important predictors for both CMP. As expected from theory, temperature, and vertical velocity are the main drivers for the prediction. Aerosol concentrations play only a minor role in the prediction. 

\begin{figure}
 \begin{subfigure}[t]{.5\textwidth}
  \centering
  \includegraphics[width=\linewidth]{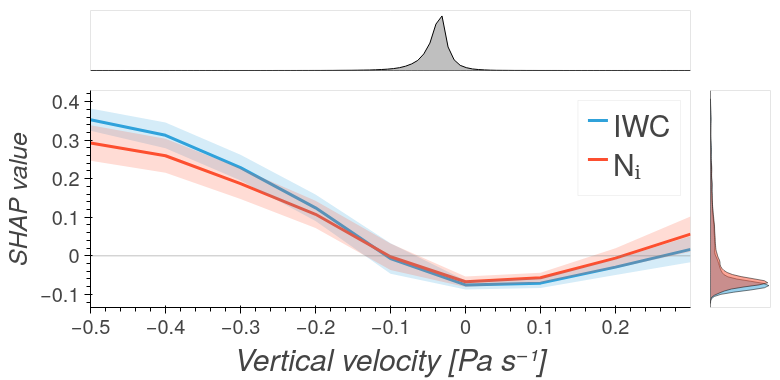}
  \caption{}
 \end{subfigure}
 \hfill
 \begin{subfigure}[t]{.5\textwidth}
  \centering
  \includegraphics[width=\linewidth]{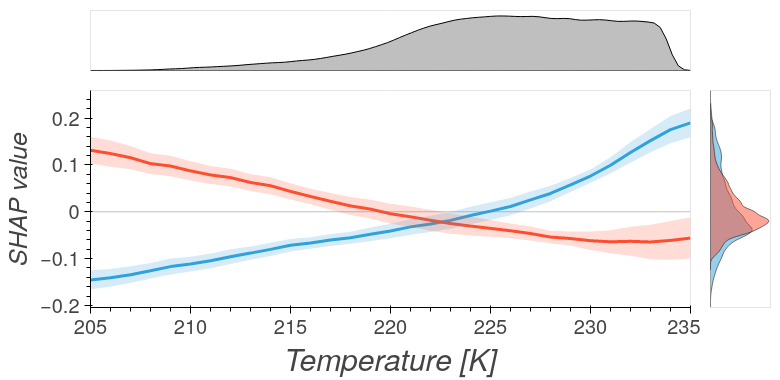}
  \caption{}
 \end{subfigure}
 \begin{subfigure}[t]{.5\textwidth}
  \centering
  \includegraphics[width=\linewidth]{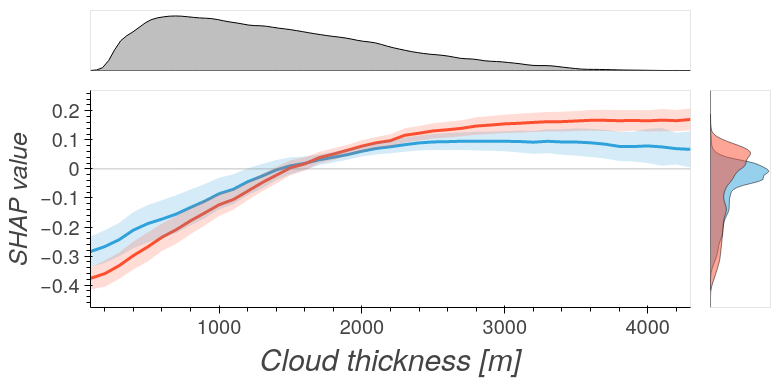}
  \caption{}
 \end{subfigure}
 \hfill
 \begin{subfigure}[t]{.5\textwidth}
  \centering
  \includegraphics[width=\linewidth]{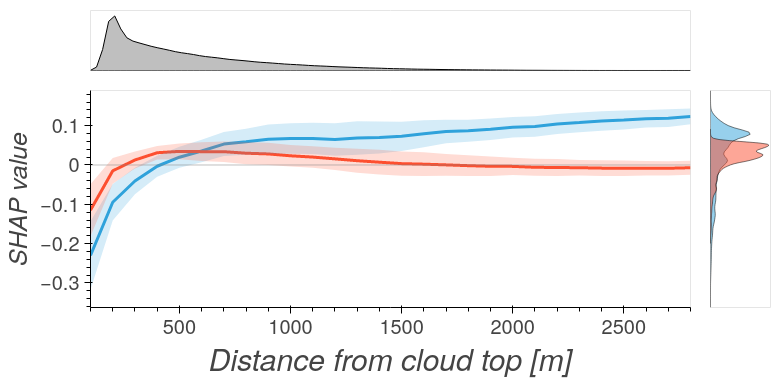}
  \caption{}
 \end{subfigure}
 \begin{subfigure}[t]{.5\textwidth}
  \centering
  \includegraphics[width=\linewidth]{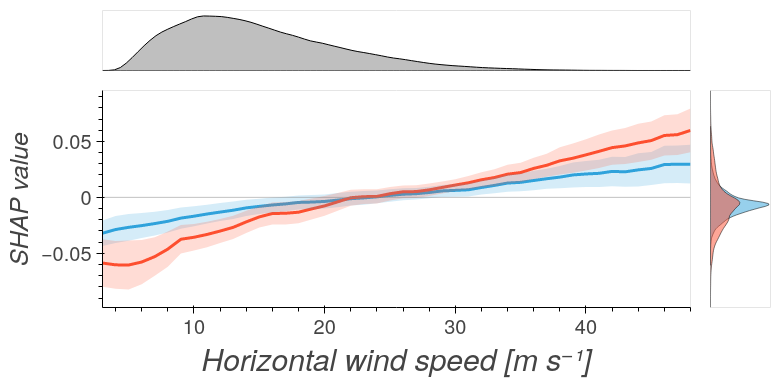}
  \caption{}
 \end{subfigure}
 \hfill
 \begin{subfigure}[t]{.5\textwidth}
  \centering
  \includegraphics[width=\linewidth]{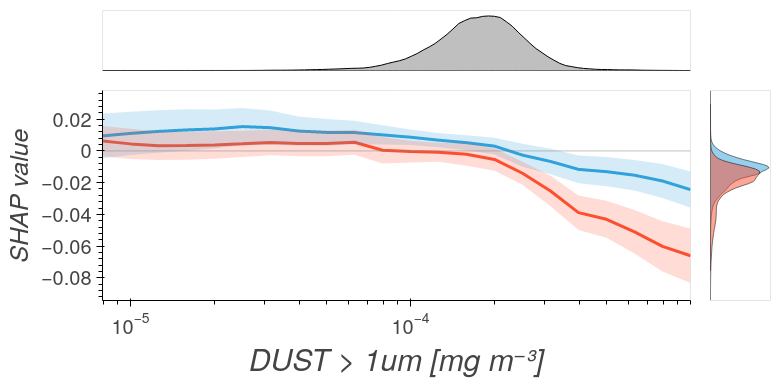}
  \caption{}
 \end{subfigure}
 \begin{subfigure}[t]{0.95\textwidth}
  \centering
  \includegraphics[width=\linewidth]{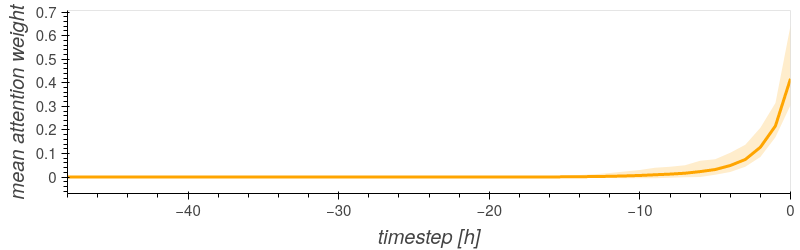}
  \caption{}
 \end{subfigure}
{\caption{\textbf{(a)-(f)}: Partial SHAP dependence plots for representative features showing the mean SHAP value per feature value for $IWC$ in blue color and $N_i$ in red color as solid lines. The shaded area represents the standard deviation. Additionally, the marginal distribution of the feature is displayed in grey above each plot, and the marginal SHAP value distributions are displayed on the right of each plot. Note that the absolute SHAP values of $IWC$ and $N_i$ are not directly comparable as the SHAP value indicates the contribution of a feature to the prediction value, and the two variables have different yet similar distributions. The plots show only feature values with at least $5000$ occurrences in the test data set. 
\textbf{(g)}: Mean of attention weight per timestep. Attention weights are learned during the training process of the LSTM model and represent the relative importance of the input data per timestep. 
}\label{fig:shap_dependence_values}}
\end{figure}

Figure \ref{fig:shap_dependence_values} allows a more detailed analysis of the SHAP values. Here, we aggregated the SHAP values for each feature along the range of occurring feature values. Increasing updrafts (i.e., negative vertical velocities \textbf{(a)}) are increasing the model predictions for $IWC$ and $N_i$. Horizontal wind speed \textbf{(e)} has a near-linear contribution to the predictions with a positive impact for wind speeds exceeding $25$ ms\textsuperscript{-1}. Horizontal wind speed is not considered much in cirrus driver literature but could be a good indicator for dynamical systems such as warm conveyor belts. For temperature \textbf{(b)}, a positive dependency is found for $IWC$ and a negative dependency for $N_i$. At temperatures between $230$ K and $235$ K, there is still a strong positive dependency for $IWC$, while the curve for $N_i$ is flattening out. This could indicate that in this temperature range, keeping all other conditions the same, the main processes are growing or colliding ice crystals instead of the formation of new ice crystals. Looking at cloud thickness, it can be observed that vertically thicker clouds increase CMP regardless of where the cloud layer is positioned in the cloud. A strong decrease in CMP is visible for cloud layers at the top of the cloud, with a more pronounced effect for $IWC$ \textbf{(d)}. This suggests that the entrainment of dry air at the cloud top is a dominant process for cirrus, causing the ice crystals to shrink and partially sublimate. Dust > 1 \textmu m slightly decreases the predictions when surpassing a concentration of $2 \times 10^{-4}$ mg m\textsuperscript{-3}, with a stronger effect for $N_i$. The physical effect acting here could be the suppression of homogeneous nucleation.

\subsection{Interpretation of attention weights}
Figure \ref{fig:shap_dependence_values} (g) displays the mean attention weight per timestep of the attention layer of the LSTM model. Compared to the post-hoc feature attribution with SHAP, the attention weights are learned during the training process in the attention layer. The weights indicate how important the model considers a given timestep for the prediction. 
Results suggest that all necessary information for cirrus prediction is encapsulated in the last $15$ hours before the observations, with more information the closer the timestep to the observation.

\section{Conclusion}\label{section4}
This work shows that ML is well suited to capture the non-linear dependencies between cirrus drivers and CMP. We applied a two-fold approach by training an XGBoost model on instantaneous cirrus driver data and an LSTM model with an attention layer on temporally resolved cirrus driver data. The LSTM-based model yielded the best predictive performance, suggesting that the history of an air parcel leading up to a cirrus cloud observation contains useful information. By analyzing the weights of the attention layer, we can show that only the last $15$ hours before the observation is of interest. Furthermore, we calculate post-hoc feature attributions for the XGBoost model using SHAP values. The analysis of SHAP values enables us to quantify the impact of driver variables on the prediction of CMP. While the physical interdependencies are mostly already known and discussed in the literature, our study can convert qualitative estimates of the dependencies to quantitative ones, such as the slope of the temperature dependency or the dust aerosol concentration needed to see an impact on the CMP. We also demonstrate the importance of vertical cloud extent and cloud layer position to the CMP, apart from meteorological and aerosol variables.
Our study is limited by the imperfect prediction of cirrus CMP, reducing the accuracy of the feature attributions. 
The authors assume that sub-grid-scale processes in reanalysis data, like small-scale updrafts and satellite retrieval uncertainties, reduce predictive performance. 
In conclusion, ML algorithms can predict cirrus properties from reanalysis and macrophysical observational data. Model explainability methods, namely SHAP and attention mechanisms, are helpful tools in improving process understanding.

\paragraph{Acknowledgments}
We are grateful for the funding from the European Union’s Horizon 2020 program. The authors would like to thank Heini Wernli, Hanin Binder, and Michael Sprenger for providing the backtrajectory data, Odran Sourdeval for helpful discussions on the DARDAR-Nice data set, and Miguel-Ángel Fernández-Torres for discussion on the ML implementation.

\paragraph{Funding Statement}
This research was supported by grants from the European Union’s Horizon 2020 research and innovation program iMIRACLI under Marie Skłodowska-Curie grant agreement No 860100 and Swiss National Supercomputing Centre (Centro Svizzero di Calcolo Scientifico, CSCS; project ID s1144)

\paragraph{Competing Interests}
The authors declare no competing interests exist.

\paragraph{Data Availability Statement}
The collocated data set and source code are available at \url{https://zenodo.org/record/7965381} and \url{https://github.com/tabularaza27/explaining_cirrus} respectively. The original data sets are freely available online: DARDAR-Nice (\url{https://doi.org/10.25326/09}), ERA5 \citep{hersbach_era5_2018}, and MERRA2 \citep{global_modeling_and_assimilation_office_merra-2_2015}.

\paragraph{Ethical Standards}
The research meets all ethical guidelines, including adherence to the legal requirements of the study country.

\paragraph{Author Contributions}
Conceptualization: K.J.; U.L.; D.N: Methodology: K.J; U.L.; G.C.V.; D.N. Data curation: K.J. Data visualization: K.J. Writing original draft: K.J. All authors approved the final submitted draft.

\bibliographystyle{apalike}
\bibliography{references.bib}

\clearpage

\begin{appendix}

\begin{center}
\textbf{\large Supplementary Material: Understanding cirrus clouds using explainable machine
learning}
\end{center}
\pagenumbering{roman}
\counterwithin{figure}{section}
\counterwithin{table}{section}

\section{ML Model Hyperparameters}\label{appendixExperimental setup}

\begin{table}[!htb]
    \small
    \caption{Best performing hyperparameters for both ML models found by Bayesian optimization.}
    \label{table:hparams}
    \begin{subtable}{.4\linewidth}
      \centering
        \caption{XGBoost}
        \begin{tabular}{ll}
            Maximum tree depth & 15 \\
            Alpha & 38 \\
            Lambda & 7 \\
            Subsample ratio of the training data & 0.4 \\
            Column subsample ratio for each tree & 0.8 \\
            Number of trees & 250 \\
            Learning rate & 0.02 \\ \\
        \end{tabular}
    \end{subtable}%
    \hspace*{\fill}
    \begin{subtable}{.55\linewidth}
      \centering
        \caption{LSTM + Attention}
        \begin{tabular}{lll}
            \multirow{1}{*}{LSTM} & hidden layer size & 250 \\ 
            \cdashline{1-3}
            \multirow{4}{*}{Final layers} & Layer sizes & 100, 50 \\
                                          & Dropout & 0.5 \\
                                          & Activation function & ReLU \\
                                          & Regularizer & Batch normalization \\
            \cdashline{1-3}
            \multirow{3}{*}{General} & Maximum epochs & 50 \\
                                     & Batch size & 1000 \\
                                     & Learning rate & 1e-5 \\
        \end{tabular}
    \end{subtable} 
\end{table}

\section{XAI evaluation metrics}\label{appendixXAIMetrics}
In this section, we provide further details on the metrics used to evaluate the post-hoc explanations. Since the metrics were originally proposed for classification tasks, they were adapted to fit the regression setting of this study. \\

\textbf{Stability} Relative Input Stability (RIS) and Relative Output Stability (ROS) are applied to evaluate the robustness of post-hoc feature attributions towards small changes in the input data. First, a modified input data set is created by adding random noise to the original samples. Then, explanations for the slightly modified samples are calculated. Finally, the relative distance between original and modified explanations with respect to the distance between the original and modified sample (RIS) and original and modified prediction (ROS) are calculated with smaller relative distances representing more stable explanations. The code to calculate the stability metrics was adapted from \citepSM{agarwal_openxai_2022}. \\

\textbf{Estimated Faithfulness} The metric is calculated by incrementally removing each of the attributes deemed important by the post-hoc feature attribution method and evaluating the effect on the performance. Features are removed by replacing them with their mean value. The correlation between the importance score of a feature and the feature's effect on model performance yields the faithfulness measure with higher correlations representing more faithful explanations. Our implementation is an adaption of \citepSM{arya_one_2019}.

\bibliographystyleSM{apalike}
\bibliographySM{references.bib}

\end{appendix}

\end{document}